\documentstyle[12pt]{article}
\begin{document}

\title{\bf Spectral statistics and periodic orbits}
\author{{\it E.B. Bogomolny}  \\
\\  
Laboratoire de Physique Th\'eorique\\
et\\
Mod\`eles Statistiques \\ 
\\
Universit\'e Paris - Sud\\ 
91405  Orsay Cedex, France }

\maketitle
\begin{abstract}
The main purpose of these lectures is to discuss briefly recent methods of
calculation of statistical properties of quantum eigenvalues for chaotic
systems based on semi-classical trace formulas. 
Under the assumption that periodic orbit actions are non-commensurable 
it is demonstrated by a few different methods that the spectral statistics of  
chaotic systems without time-reversal invariance in the universal limit 
agrees with statistics of the Gaussian Unitary Ensemble of random matrices. 
The methods used permit to obtain not only the limiting statistics but also
the way the spectral statistics of dynamical systems tends to the universal limit.
The statistics of the  Riemann zeta function zeros is considered in details.
\end{abstract}

\pagebreak

\section{Generalities}

\subsection{Trace formulas}

The quantum density of states for a chaotic dynamical systems 
\begin{equation}
d(E)=\sum_n \delta(E-E_n)
\end{equation}
under quite general conditions can be expressed through a sum over classical 
periodic orbits by  the {\bf Gutzwiller trace formula} \cite{Gutzwiller1},
\cite{Gutzwiller2} (plus corrections if necessary)
\begin{equation}
d(E)=\bar{d}(E)+d^{(osc)}(E),
\end{equation}
where the smooth part, $\bar{d}(E)$, for a $f$-dimensional system
with a Hamiltonian $H(\vec{p},\vec{q}\ )$ is given by the Thomas-Fermi formula
\begin{equation}
\bar{d}(E)=\int \frac{d\vec{p}d\vec{q}}{(2\pi \hbar)^f}
\delta (E-H(\vec{p},\vec{q}\ )),
\end{equation}
and the (most interesting) oscillating part of the level density,
$d^{(osc)}(E)$, has the following form 
\begin{equation}
d^{(osc)}(E)=\frac{1}{\pi \hbar}\sum_{ppo}T_p\sum_{n=1}^{\infty}
\frac{1}{|\det (M_p^n-1)|^{1/2}}\cos(n(\frac{S_p}{\hbar}-\frac{\pi}{2}\mu_p)).
\label{4}
\end{equation}
Here the first summation is done over primitive classical periodic orbits (ppo)
with enrgy $E$ and the
second one is performed over all repetitions of a given ppo.
$S_p$ is the classical action calculated for a ppo labeled by $p$, $M_p$ is
the monodromy matrix for this trajectory, and $\mu_p$ is its Maslov index
\cite{Gutzwiller1}, \cite{Gutzwiller2}.

Similar formulas can be written also for other spectral functions.\\
In particular the {\bf staircase function}  has the following form
\begin{equation}  
N(E)=\int^{E}_{-\infty}d(E')dE'=\bar{N}(E)+N^{(osc)}(E),
\end{equation}
where
\begin{equation}
\bar{N}(E)=\int \frac{d\vec{p}d\vec{q}}{(2\pi \hbar)^f}
\Theta (E-H(\vec{p},\vec{q})),
\end{equation}
and 
\begin{equation}
N^{(osc)}(E)=\frac{1}{\pi}\sum_{ppo}\sum_{n=1}^{\infty}
\frac{1}{n |\det (M_p^n-1)|^{1/2}}\sin(n(\frac{S_p}{\hbar}-\frac{\pi}{2}\mu_p)).
\end{equation}
We shall need also the {\bf  dynamical zeta function} which  (for
2-dimensional systems)  equals the following product over ppo 
\begin{equation}
Z(E)=\prod_{ppo}\prod_{m=0}^{\infty}(1-\frac{e^{iS_p/\hbar-i\pi \mu_p/2}}
{|\Lambda_p |^{1/2}\Lambda_p^m}).
\label{8}
\end{equation}
Here $\Lambda_p$ is the largest eigenvalue of the monodromy matrix
($|\Lambda_p|>1$)
\begin{equation}
M_p u=\Lambda_p u.
\end{equation}
This zeta function is connected with the density of states and the
staircase function by the following relations
\begin{equation}
d^{(osc)}(E)=-\frac{1}{2\pi i}( \frac{Z'(E)}{Z(E)}-c.c.),
\end{equation}
and
\begin{equation}
N^{(osc)}(E)=-\frac{1}{2\pi i}( \ln Z(E)-c.c.).
\end{equation}
The trace formulas exist not only for dynamical systems but also for 
the {\bf Riemann zeta function} \cite{Titchmarsh} (and others
number-theoretical zeta functions as well). 

The Riemann zeta function is defined as 
\begin{equation}
\zeta (s)=\sum_{n=1}^{\infty}\frac{1}{n^s}=\prod_p \frac{1}{(1-p^{-s})},
\end{equation}
where the product is taken over prime numbers. This function converges only
when Re$s>1$ but it is well known \cite{Titchmarsh} that it can be
analytically  continued in the whole complex $s$-plane, the only
singularity being the pole at $s=1$ with unit  residue.

According to the famous Riemann conjecture all nontrivial zeros of this
function have the form $s_n=\frac{1}{2}+iE_n$ and the
density of these zeros $d(E)=\sum_n\delta (E-E_n)$ can be expressed by the
following `trace' formula
\begin{equation}
d(E)=\bar{d}(E)+d^{(osc)}(E),
\end{equation}
where
\begin{equation}
\bar{d}(E)=\frac{1}{2\pi}\ln \frac{E}{2\pi},
\end{equation}
and 
\begin{equation}
d^{(osc)}(E)=-\frac{1}{\pi}\sum_p\sum_{n=1}^{\infty} \frac{\ln p}{p^{n/2}}
\cos(En\ln p),
\label{16}
\end{equation}
where the summation is performed over all prime numbers.

The staircase function for the Riemann zeros has a similar form
\begin{equation}
N(E)=\bar{N}(E)+N^{(osc)}(E),
\end{equation}
where
\begin{equation}
\bar{N}(E)=\frac{E}{2\pi}(\ln \frac{E}{2\pi}-1)+\frac{7}{8},
\label{17}
\end{equation}
and
\begin{equation}
N^{(osc)}(E)=-\frac{1}{\pi}\sum_p\sum_{n=1}^{\infty} \frac{1}{np^{n/2}}
\sin(En\ln p).
\end{equation}
By comparing Eq.~(\ref{4}) and Eq.~(\ref{16}) one observes a remarkable
correspondence between different quantities in these trace formulas
\cite{Berry}
\begin{center}
Periodic orbits of chaotic systems $\leftrightarrow $ primes,\\
periodic orbit period $\leftrightarrow $ $\ln p$.
\end{center}
The convergence properties of both formulas are also quite similar. The
number of periodic orbits with period less than $T$ for chaotic systems is
asymptotically (see e.g. \cite{Lichtenberg})
\begin{equation}
N(T_p<T)=\frac{e^{hT}}{hT},
\label{20}
\end{equation}
where constant $h$ is called the topological entropy. The number of prime
numbers less than $x$ is given by the prime number theorem \cite{Titchmarsh} 
\begin{equation}
N(p<x)\equiv \pi(x)=\frac{x}{\ln x}.
\end{equation}
As $\ln p \equiv T_p$ this expression has the form similar to
(\ref{20}) with $h=1$
\begin{equation}
N(T_p<T)=\frac{e^{T}}{T}.
\end{equation}
Due to these similarities number-theoretical  zeta functions play the role
of simple (but non-trivial) models of quantum chaos and in
these lectures we  consider in parallel both dynamical systems and zeta
functions and shall check our methods first on the Riemann case. 

\subsection{Random matrix theory}

Wigner and Dyson in the fifties had proposed to describe complicated 
(and mostly unknown) Hamiltonian of heavy nuclei by a member of an ensemble of
random matrices (see \cite{Porter})  and they argued that  the type of this
ensemble  depends only on the symmetry of the Hamiltonian. 

For systems without time-reversal invariance the relevant ensemble is the
Gaussian Unitary Ensemble (GUE), for systems invariant with respect to 
time-reversal the  ensemble should be the Gaussian Orthogonal Ensemble (GOE) 
and for systems with time-reversal invariance but with half-integer spin
energy levels should be described according to the Gaussian Symplectic
Ensemble (GSE) of random matrices \cite{Porter, Mehta}. For these
classical ensembles all correlation functions can be written explicitly
\cite{Mehta}. 
In particular {\bf 2-point correlation functions}
\begin{equation}
R_2(x)=1+\delta (x) -Y_2(x),
\end{equation}
have the following form \cite{Mehta}, \cite{Bohigas}.

For GOE
\begin{equation}
Y_2(x)=(\frac{\sin \pi x}{\pi x})^2-(\mbox{Si}(\pi x)-\frac{\pi}{2} \epsilon(x))
(\frac{\cos \pi x}{\pi x}-\frac{\sin \pi x}{(\pi x)^2}).
\end{equation}
Here
\begin{equation}
\mbox{Si} (x)=\int_0^x\frac{\sin y}{y}dy,
\end{equation}
and $\epsilon(x)=\mbox{sgn} (x)$.

For GUE
\begin{equation}
Y_2(x)=(\frac{\sin \pi x}{\pi x})^2.
\label{25}
\end{equation}

For GSE
\begin{equation}
Y_2(x)=(\frac{\sin 2\pi x}{2\pi x})^2-\mbox{Si}(2\pi x)
(\frac{\cos 2\pi x}{2\pi x}-\frac{\sin 2\pi x}{(2\pi x)^2}).
\end{equation}
{\bf 2-point correlation form factor} is the Fourier transform of the 2-point
correlation function 
\begin{equation}
K(t)=\int_{-\infty}^{\infty}R_2(x)e^{2\pi i  t x}dx,
\label{27}
\end{equation}
and for three classical ensembles it has the following forms.

For GOE
\begin{equation}
K(t)=\left \{ \begin{array}{lc}
2t-t\ln (1+2t) & 0<t<1\\
2-t\ln((2t+1)/(2t-1))& t>1
\end{array}\right . .
\label{28}
\end{equation}
 
 For GUE
\begin{equation}
K(t)=\left \{ \begin{array}{lc}
t & 0<t<1\\
1 & t>1
\end{array}\right . .
\label{29}
\end{equation}

For GSE
\begin{equation}
K(t)=\left \{ \begin{array}{lc}
\frac{1}{2}t-\frac{1}{4}t\ln (|1-t|) & 0<t<2\\
1 & t>2
\end{array}\right . .
\end{equation}
The nearest neighbor distributions for classical ensembles can be expressed
through solutions of  certain integral equations and numerically they are
close to the Wigner surmises
\begin{eqnarray}
\mbox{GOE :}& & p(s)=\frac{\pi}{2}s\exp (-\frac{\pi}{4}s^2).\\
\mbox{GUE :}& & p(s)=\frac{32}{\pi^2}s^2\exp (-\frac{4}{\pi}s^2).\\
\mbox{GSE :}& & p(s)=\frac{2^{18}}{3^6\pi^3}s^4\exp (-\frac{64}{9\pi}s^2).
\end{eqnarray}

Later it was understood that the same conjectures can be applied not only
for heavy nuclei but also for simple dynamical systems and to-day 
{\bf standard accepted conjectures} are the following

\begin{itemize}
\item The energy levels of classically integrable systems on the scale of
  the mean level density behave as independent random variables and their
  distribution is close to the Poisson distribution (Berry, Tabor 
  \cite{BerryTabor}).
\item The energy levels of classically chaotic systems are not independent
  but on the scale of the mean level density they are distributed as 
  eigenvalues of random matrix ensembles depending
  only on symmetry properties of the system considered (Bohigas, Giannoni,
  Schmit \cite{BohGianSchmit}). 
\begin{itemize}
\item  
  For systems without time-reversal invariance the distribution of energy
  levels should be close to the distribution of the Gaussian Unitary
  Ensemble (GUE) characterized by quadratic level repulsion. 
\item
  For systems with time-reversal invariance the corresponding distribution
  should be close to that of the Gaussian Orthogonal Ensemble (GOE) with
  linear level repulsion.
\item
  And for systems with time-reversal invariance but with half-integer spin
  energy levels should be described according to the Gaussian Symplectic
  Ensemble (GSE) of random matrices with quartic level repulsion.
\end{itemize}  
\end{itemize}
These conjectures are very well confirmed by numerical calculations (see
e.g. \cite{Bohigas}, \cite{Weidenmuller}). The main purpose of these lectures
is the developing of methods which permit to prove analytically one part of these
conjectures, namely that in the universal limit (i.e. on the scale of
  the mean level density) the 2-point correlation
function  for systems without time-reversal invariance agrees with
Eq.~(\ref{25}). For further references it is convenient to rewrite this
expression in the dimensional form
\begin{equation}
\tilde{R}_2(\epsilon)=\bar{d}^2R_2(\bar{d}\epsilon).
\end{equation}
It gives
\begin{equation}
\tilde{R}_2(\epsilon)=\bar{d}^2+\bar{R}_2(\epsilon)+R_2^{(osc)}(\epsilon),
\end{equation}
where the smooth part 
\begin{equation}
\bar{R}_2(\epsilon)=-\frac{1}{2\pi^2 \epsilon^2},
\label{25a}
\end{equation}
and the oscillating part
\begin{equation}
R_2^{(osc)}(\epsilon)=
\frac{e^{2\pi i \bar{d}\epsilon}+e^{-2\pi i \bar{d}\epsilon}}
{4\pi^2 \epsilon^2}.
\label{25b}
\end{equation}
The plan of the paper is the following. In Section \ref{correlation} we
relate through the trace formulas correlation functions of quantum eigenvalues
(and of Riemann zeros) with periodic orbit sums and in Section \ref{diagonal} 
the simplest approximation of evaluating these sums called the diagonal
approximation is discussed. But as shown in Section \ref{limit} this
approximation is
valid only for large energy difference and  to find correlation
functions in the full range new methods are needed. In Section \ref{beyond} 
a  method specific for the Riemann zeta function is discussed. To calculate
off-diagonal terms we use the Hardy-Littlewood conjecture for distribution
of pairs of primes and in Section \ref{Hardy} it is demonstrated that this
conjecture leads to  interesting formulas for the 2-point
correlation function of the Riemann zeros. In the universal limit this
correlation function tends to the GUE results but it also permits to
investigate how the correlation function tends to the universal result. 
In Section \ref{arithmetical} we briefly mention a certain model where a
generalization of the Hardy-Littlewood conjecture also permits to calculate
the 2-point correlation function. But the Hardy-Littlewood conjecture can be
aplied only for primes and it cannot be generalized for dynamical systems.
In Section \ref{finite} a certain method is proposed which overcomes this
difficulty. The main ingredient of this method is an artificial construction
of approximate density of states which has the correct analytical properties
but requires the knowledge only of finite number of periodic orbits. In
Section \ref{offriemann} it is shown that for the case of the Riemann zeros
the method gives exactly the same result as was obtained by using the
Hardy-Littlewood conjecture. In Section \ref{offdynamical} this method is
applied to dynamical systems and it was demonstrated that under the
assumption that all periodic orbit actions are non-commensurable it is
possible to prove that  spectral statistics of generic dynamical systems
without time-reversal invariance in the universal limit tends to the GUE
statistics. In Section \ref{universal} an other method of calculation of
spectral statistics is proposed. The method is based on the universality of
different random matrix ensembles and it is shown that it gives exactly the
same expressions for 2-point correlation functions for the Riemann zeros and
dynamical systems as have been obtained in previous Sections by different
methods. Finally in Section \ref{RiemannSiegel} one more method is discussed
which is based on the Riemann-Siegel type  resummation of the trace formulas
which also leads to the same expressions for  2-point correlation functions. 
Though the exact mathematical proof of the above results is not known and
all our methods should be considered as heuristics or `physical' proofs 
the agreement between different methods strongly indicates the correctness
of the results.

\section{Correlation functions}\label{correlation}

Formally 
$n$-point correlation functions of energy levels are defined as the
probability of having $n$ energy levels at given positions and they are
connected to the density of states by the  relations
\begin{equation}
R_n(\epsilon_1,\epsilon_2,\ldots,\epsilon_n)=
<d(E+\epsilon_1)d(E+\epsilon_2)\ldots d(E+\epsilon_n)>,
\label{34}
\end{equation}
where the brackets $<\dots>$ denote the smoothing over an appropriate energy
window
\begin{equation}
<f(E)>=\int f(E')\sigma (E-E')dE' ,
\end{equation}
with
\begin{equation}
\int \sigma(E)dE=1.
\end{equation}
Function $\sigma(E)$ is assumed centered near zero and has a width
$\Delta E$ which obeys inequalities
\begin{equation}
\Delta E_q \ll \Delta E \ll \Delta E_{cl} \ll E.  
\end{equation}
Here $\Delta E_q $ is of the order of the mean level separation,
$\Delta E_q \approx 1/\bar{d}$, and
$\Delta E_{cl} $ denotes the energy scale at which classical dynamics
changes noticeable.

Let us write the trace formula for the density of states (\ref{4}) in the form
\begin{equation}
d(E)=\bar{d}(E)+\sum_{p,n}T_p A_{p,n}e^{i n S_p(E)/\hbar} +c.c.,
\end{equation}
where
\begin{equation}
A_{p,n}=\frac{1}{2\pi\hbar|\det (M_p^n-1)|^{1/2}}e^{-\pi in\mu_p/2}.
\end{equation}
Substituting this expression to the formula for 2-point correlation
function one gets  
\begin{eqnarray}
&&R_2(\epsilon_1, \epsilon_2)=\bar{d}^2+\\
&&\sum_{p_1,p_2}T_{p_1}T_{p_2}A_{p_1,n_1}A_{p_2,n_2}^*
<\exp \frac{i}{\hbar}(S_{p_1}(E+\epsilon_1)-S_{p_2}(E+\epsilon_2))> +c.c.,
\nonumber
\end{eqnarray}
and the terms with the sum of actions are assumed to be washed out by the 
smoothing procedure.

Expanding the actions and taking into account that $\partial S(E)/\partial E=T(E)$
where $T(E)$ is the classical period of motion one finds
\begin{eqnarray}
R_2(\epsilon_1, \epsilon_2)&=&\bar{d}^2+
\sum_{p_i,n_i}A_{p_1,n_1}A_{p_2,n_2}^*
<\exp \frac{i}{\hbar}(n_1 S_{p_1}(E)-n_2 S_{p_2}(E))>
\nonumber\\
&& \times 
\exp \frac{i}{\hbar}(n_1 T_{p_1}(E)\epsilon_1-n_2 T_{p_2}(E)\epsilon_2)+c.c.
\end{eqnarray}

\subsection{\bf Diagonal approximation}\label{diagonal}

Berry \cite{Berry}
proposed to estimate this sum by taking into account terms only with
exactly the same actions having in mind that terms with different values of
actions should be small after the smoothing.
Therefore in this approximation (called the diagonal approximation)
\begin{equation}
R_2^{(diag)}(\epsilon)=\sum_{p,n}T_p^2|A_{p,n}|^2e^{i n T_p(E)\epsilon} +c.c.,
\end{equation}
Here $\epsilon=\epsilon_1-\epsilon_2$ and the sum is taken over all periodic 
orbits with exactly the same action.

Introducing the classical zeta function
\footnote{Often the classical zeta function is introduced in such a manner that
$\sum_{ppo}\sum_{k=1}^{\infty}\frac{e^{sT_p k}}{k|\det (M_p^k-1)|}=\ln Z_{cl},$
which gives
$Z_{cl}(s)=\prod_{ppo}\prod_{m=0}^{\infty}
(1-\frac{e^{sT_p}}{|\Lambda_p|\Lambda_p^m})^{-(m+1)}.$
As the convergent properties of both zeta functions are the same we chose the
simplest definition sufficient for our purposes.}\label{footnote}
\begin{equation}
Z_{cl}(s)=\prod_{ppo}(1-\frac{e^{sT_p}}{|\Lambda_p|})^{-1}
\end{equation}
one can rewrite the 2-point correlation function in the form
\begin{equation}
R_2^{(diag)}(\epsilon)=-\frac{1}{4\pi^2}\frac{\partial^2}{\partial \epsilon^2}
\ln \Delta (\epsilon),
\end{equation}
where
\begin{equation}
\Delta (\epsilon)=|Z_{cl}(i\epsilon)|^2\Phi^{(diag)}(\epsilon),
\end{equation}
and function $\Phi^{(diag)}(\epsilon)$ is defined as the following 
convergent sum over periodic orbits
\begin{equation}
\Phi^{(diag)}(\epsilon)=\exp \left (\sum_{ppo}\sum_{m=1}^{\infty}\frac{1}{m|\Lambda_p|^m}
(\frac{1}{m(1-\Lambda_p^{-m})^2}-1)e^{i m T_p(E)} + c.c. \right ).
\label{40}  
\end{equation}
Another useful form of these relations is the expression for the diagonal
approximation of 2-point form factor defined in Eq.~(\ref{27})
\begin{equation}
K^{(diag)}(t)= 2\pi \sum_{p,n}T_p^2|A_{p,n}|^2\delta (2\pi t-nT_p(E)) +c.c.
\end{equation}
According to the Hannay-Ozorio de Almeida sum rule for ergodic systems 
\cite{HannayOzorio} one can compute such sums by substituting the local
density of periodic orbits 
\begin{equation}
\rho_p=\frac{|\det (M_p-1)|}{T_p},
\end{equation}
and  consequently
\begin{equation}
K^{(diag)}(t)=\frac{g}{2\pi} \int T_p\delta (2\pi t-T_p)dT_p=gt,
\label{49}
\end{equation}
where $g$ is the mean multiplicity of periodic orbits (i.e. the number of
periodic orbits with exactly the same action). For generic systems
without time-reversal invariance there is no reasons for equality of actions
for different periodic orbits and $g=1$ but for systems with time-reversal
invariance each orbits can be passed in two directions therefore  in general 
for such systems $g=2$. Comparing (\ref{49}) with Eqs.~(\ref{28}) and
(\ref{29}) one concludes that the diagonal approximation reproduces the correct  
small-$t$ behavior of form-factors of classical ensembles. 

Unfortunately, $K^{(diag)}(t)$ grows with increasing of $t$ but the exact
form-factor for systems without spectral degeneracy should tends to $\bar{d}$
for large $t$ which reflects the existence of the delta function in
$R_2(\epsilon)$ for small $\epsilon$ \cite{Berry}
\begin{equation}
R_2(\epsilon)\rightarrow \bar{d}\delta(\epsilon), \;\;\mbox{ when }
\epsilon \rightarrow 0,
\end{equation}
or
\begin{equation}
K(t)\rightarrow \bar{d}, \;\;\mbox{ when }t\rightarrow \infty.
\end{equation}
This evident contradiction clearly indicates that the diagonal approximation
cannot be correct for all values of $t$ and more complicated tools are needed
to obtain the full form-factor.

\subsection{\bf Criterion of applicability of diagonal
  approximation}\label{limit}

One can give a (pessimistic) estimate till what time the diagonal
approximation can be correct by the following method. The main ingredient of the
diagonal approximation is the assumption that after smoothing all
off-diagonal terms give negligible contribution. This condition is almost the same
as the condition of the absence of quantum interference. But it is known that
the quantum interference is not important for times smaller than the
Ehrenfest time which is of the order of
\begin{equation}
t_E\approx \frac{1}{\lambda}\ln (1/\hbar),
\end{equation}
where $\lambda$ is a (classical) constant of the order of the Lyapunov
exponent defined in such a way that the mean splitting of two nearby
trajectories for the
time $t$ grows as $\exp (\lambda t)$. For billiards $\lambda=k\lambda'$
where  $k$ is the momentum and $\lambda'$ determines the deviation of two 
trajectories with the length $L=kt$, and the role of $\hbar$ plays  $k^{-1}$
\begin{equation}
t_E\approx \frac{1}{\lambda'k}\ln(k).
\end{equation}
More refined estimates can be done as follows. The off-diagonal terms can be
neglected if 
\begin{equation}
<\exp \frac{i}{\hbar}(S_{p_1}(E)-S_{p_2}(E))>\ll 1. 
\end{equation}
But this quantity is small provided the difference of period of two orbits
$\Delta T=T_{p1}-T_{p_2}$ times the energy window $\Delta E$ used in the 
definition of smoothing procedure is large
\begin{equation}
\frac{1}{\hbar}(T_{p1}-T_{p_2})\Delta E \gg 1.
\end{equation}
For billiards $T_p=L_p/k$ and this condition means that one has to consider
all periodic orbits such that their difference of lengths is 
\begin{equation}
L_{p1}-L_{p_2}\gg \frac{\hbar k}{\Delta E}.
\end{equation}
But the number of periodic orbits with the length $L$ grows exponentially
\begin{equation}
N(L_p<L)=\frac{e^{hL}}{hL},
\end{equation}
where $h$ is of the order of the Lyapunov exponent $\lambda'$.
Therefore in the interval $L, L+\delta l$ there is $e^{ hL}\delta l/L$
orbits and  the mean difference of lengths between orbits with the length
$<L$ is of the order of 
\begin{equation}
\Delta L =L\exp (-hL).
\end{equation}
To fulfilled the above condition one has to restrict the maximum length of
periodic orbits, $L_m$, by
\begin{equation}
L_m\exp (-hL_m)\approx \frac{k \hbar}{ \Delta E}
\end{equation}
In the limit of large $L_m$ it gives
\begin{equation}
L_m\approx \frac{1}{h}\ln \frac{ \Delta E}{k\hbar h},
\end{equation}
which corresponds to the Ehrenfest estimate above. 

Let us denote
$h=1/l_0$ where $l_0$ has the dimensionality of the length. Then
the above estimate can be transform into the following form
\begin{equation}
L_m\approx l_0\ln \frac{ \Delta E}{E_T},
\end{equation}
and $E_T$ is an analog of the Thouless energy
\begin{equation}
E_T=\frac{\hbar}{\tau_T}\;\;\;\mbox{and}\;\tau_T=\frac{l_0}{k}.
\end{equation}
Note that the Heisenberg time is 
\begin{equation}
t_H=2\pi \bar{d},
\end{equation}
and for billiards because $\bar{d}$ is a constant 
\begin{equation}
t_E\ll t_H.
\end{equation}

For the Riemann zeta function the situation is better because the role of
`energy' in this case plays the `momentum' and the density of states
for the Riemann zeta function is  $(\ln (E/2\pi))/(2\pi)$. As
in this case $h=1$
\begin{equation}
t_E= t_H,
\end{equation}
and the diagonal approximation for the Riemann zeta function is valid till
the Heisenberg time \cite{Montgomery}, \cite{Sarnak}.

This type of estimates directly indicates that the diagonal approximation
for dynamical systems can not, strictly speaking,  be used to obtain an
information about the form-factor for large value of $t$. (Note that for GUE
systems the diagonal approximation gives the expected answer till the
Heisenberg time but it just signifies that one has to find special
reasons why all other terms cancel.)

\section{Beyond the diagonal approximation}\label{beyond}

The simplest and the most natural way of semi-classical computation of 
2-point correlation functions is to find a method of calculating 
off-diagonal terms.
We shall discuss here this type of  computation on the example of the
Riemann zeta function where much more information than for dynamical systems 
is available.

The trace formula for the Riemann zeta function may be rewritten in the
form
\begin{equation}
d^{(osc)}(E)=-\frac{1}{\pi}\sum_{n=1}^{\infty} \frac{1}{\sqrt{n}}
\Lambda(n)\cos(E\ln n),
\end{equation}
where 
\begin{equation}
\Lambda(n)=\left \{ \begin{array}{lr}
\ln p, & \mbox{if } n=p^k\\
0, & \mbox{otherwise}
\end{array} .\right .
\end{equation}
The connected 2-point correlation function of the Riemann zeros,
$R_2^{(c)}=R_2-\bar{d}^2$, is
\begin{equation}
R_2^{(c)}(\epsilon_1,\epsilon_2)=\frac{1}{4\pi^2}\sum_{n_1,n_2}
\frac{\Lambda(n_1)\Lambda(n_2)}{\sqrt{n_1n_2}}
<e^{i(E+\epsilon_1)\ln n_1-i(E+\epsilon_2)\ln n_2}> +c.c.\;.
\end{equation}
The diagonal approximation corresponds to taking into account terms 
with $n_1=n_2$
\begin{eqnarray}
R_2^{(diag)}(\epsilon_1,\epsilon_2)&=&\frac{1}{4\pi^2}\sum_{n}
\frac{\Lambda^2(n)}{n}
(e^{i(\epsilon_1-\epsilon_2)\ln n} +c.c.)
\nonumber \\
&=&
\frac{1}{4\pi^2}\sum_{p,m}
\frac{\ln^2 p}{p^{m}}
(e^{i(\epsilon_1-\epsilon_2)m\ln p} +c.c.).
\label{69}
\end{eqnarray}
This expression may  be transform as follows
\begin{equation}
R_2^{(diag)}(\epsilon)=-\frac{1}{4\pi^2}
\frac{\partial^2}{\partial \epsilon ^2}\ln \Delta(\epsilon),
\label{70}
\end{equation}
where 
\begin{equation}
\Delta(\epsilon)=|\zeta(1+i\epsilon)|^2\Phi^{(diag)}(\epsilon),
\label{71}
\end{equation}
and function $\Phi^{(diag)}(\epsilon)$ is given by a convergent sum over
prime numbers
\begin{equation}
\Phi^{(diag)}(\epsilon)=\exp (\sum_p\sum_{m=1}^{\infty} 
\frac{1-m}{m^2p^m}e^{im\ln p \epsilon}+c.c.).
\label{72}
\end{equation}
In the limit $\epsilon \rightarrow 0$
$\zeta(1+i\epsilon)\rightarrow (i\epsilon)^{-1}$ and
$\Phi^{(diag)}(\epsilon)\rightarrow$ const. Therefore in this limit
\begin{equation}
R_2^{(diag)}(\epsilon)\rightarrow -\frac{1}{2\pi^2\epsilon^2},
\end{equation}
which agrees with the smooth part of the GUE result (\ref{25a}).

The off-diagonal contribution  has the form
\begin{equation}
R_2^{(off)}(\epsilon_1,\epsilon_2)=\frac{1}{4\pi^2}\sum_{n_1\neq n_2}
\frac{\Lambda(n_1)\Lambda(n_2)}{\sqrt{n_1n_2}}
(<e^{iE\ln (n_1/n_2)+i(\epsilon_1 \ln n_1-\epsilon_2\ln n_2)}> +c.c.).
\end{equation}
The term $\exp (iE\ln (n_1/n_2))$ oscillates quickly if $n_1$ is not
close to $n_2$. Denoting
\begin{equation}
n_1=n_2+d
\end{equation}
and expanding all smooth functions on $d$ one gets
\begin{equation}
R_2^{(off)}(\epsilon)=\frac{1}{4\pi^2}\sum_{n,d}
\frac{\Lambda(n)\Lambda(n+d)}{n}
(<e^{iE\frac{d}{n}+i\epsilon \ln n}> +c.c.),
\end{equation}
where $\epsilon=\epsilon_1-\epsilon_2$.

The main problem is clearly seen here. The function
\begin{equation}
F(n,d)=\Lambda(n)\Lambda(n+d)
\end{equation}
is quite a wild function as it is nonzero only when both $n$ and $n+d$ are
power of prime numbers. As we have assumed that $d\ll n$ it is quite natural
to assume that the dominant contribution to the 2-point correlation
function will come from the mean value of this function over all $n$, i.e.
one has to substitute into $R_2^{(off)}(\epsilon )$ instead of $F(n,d)$ its mean value
\begin{equation}
\alpha (d)=\lim_{N\rightarrow \infty}\frac{1}{N}\sum_{n=1}^N\Lambda(n)\Lambda(n+d).
\end{equation}

\subsection{\bf The Hardy-Littlewood conjecture}\label{Hardy}

Fortunately the explicit expression for this function comes from the famous
Hardy--Littlewood conjecture \cite{HardyLittlewood}. There are two different forms
of this conjecture which, of course, can be mutually transformed
\begin{equation}
\alpha (d)=C_2\prod_{p|d}\frac{p-1}{p-2},
\end{equation}
where the product is taken over all prime divisors of $d$ bigger than 2 and
$C_2$ is the so-called twin prime constant
\begin{equation}
C_2=2\prod_{p>2}(1-\frac{1}{(p-1)^2})\approx 1.32032\ldots .
\end{equation}
The other form which will be useful for us is expressed through the
so-called singular series 
\begin{equation}
\alpha(d)= \sum_{(p,q)=1}\exp (2\pi i \frac{p}{q}d)
\left (\frac{\mu(q)}{\psi(q)}\right )^2,
\end{equation}
where the sum is taken over all natural $q$ and all integer $p$ co-prime to
$q$ $ (p<q)$. Function $\mu(n)$ is the Mobius function defined through the
factorization of $n$ on prime factors
\begin{equation}
\mu(n)=\left \{ \begin{array}{cl}
1& \mbox{if } n=1\\
(-1)^k& \mbox{if } n=p_1\ldots p_k\\
0&\mbox{if $n$ is divisible on $p^2$}
\end{array}.\right .  
\end{equation}
Function $\psi(n)$ is the Euler function which counts the number of integers
smaller than $n$ and co-prime to $n$
\begin{equation}
\psi(n)=n\prod_{p|n}(1-\frac{1}{p}),
\end{equation}
where the product is taken over all prime divisors of $n$.

Taking  the above formulae as granted we get
\begin{equation}
R_2^{(off)}(\epsilon)=\frac{1}{4\pi^2}\sum_n
\frac{1}{n}e^{i\epsilon \ln n}
\sum_d \alpha(d)e^{iE\frac{d}{n}} +c.c.
\end{equation}
After substitution the formula for $\alpha(d)$ and performing the sum over
all $d$ one obtains
\begin{equation}
R_2^{(off)}(\epsilon)=\frac{1}{4\pi^2}\sum_n
\frac{1}{n}e^{i\epsilon \ln n}
\sum_{(p,q)=1}\left (\frac{\mu(q)}{\psi(q)}\right )^2
\delta(\frac{p}{q}-\frac{E}{2\pi n}) +c.c.,
\end{equation}
where the summation is taken over all pairs of mutually co-prime positive
integers $p$ and $q$ (without the restriction $p<q$).

Changing the sum over $n$ to the integral permits to transform  this
expression to the sum over that values of $n$ where
$$\frac{p}{q}-\frac{E}{2\pi n}=0,$$
and in this approximation
\begin{equation}
R_2^{(off)}(\epsilon)=\frac{1}{4\pi^2}e^{i\epsilon \ln \frac{E}{2\pi}}
\sum_{(p,q)=1}\left (\frac{\mu(q)}{\psi(q)}\right )^2
(\frac{q}{p})^{1+i\epsilon} +c.c.
\end{equation}
Using the formula
\begin{equation}
\sum_{(p,q)=1}f(p)=\sum_{k=1}^{\infty}\sum_{\delta |q}f(k\delta)\mu(\delta),
\end{equation}
and taking into account that $2\pi \bar{d}=\ln (E/2\pi)$ we obtain
\begin{equation}
R_2^{(off)}(\epsilon)= \frac{1}{4\pi^2}|\zeta(1+i\epsilon)|^2
e^{2\pi i\bar{d}\epsilon }\Phi^{(off)}(\epsilon)+c.c.,
\end{equation}
where function $\Phi^{(off)}(\epsilon)$ is given by a convergent product over primes
\begin{equation}
\Phi^{(off)}(\epsilon)= \prod_p(1-\frac{(1-p^{i\epsilon})^2}{(p-1)^2}),  
\end{equation}
and $\Phi^{(off)}(0)=1$.

In the limit of small $\epsilon$
\begin{equation}
R_2^{(off)}(\epsilon)= \frac{1}{(2 \pi \epsilon)^2}
(e^{2\pi i\bar{d}\epsilon }+e^{-2\pi i\bar{d}\epsilon }),
\end{equation}
which exactly corresponds to the GUE results for the oscillating part of the
2-point correlation function (\ref{25b}).

The above calculations demonstrates how one can compute the 2-point
correlation function through the knowledge of pair-correlation function of
periodic orbits. For the Riemann case one can prove under the same
conjectures that all $n$-point correlation functions of Riemann zeros tend
to corresponding GUE results \cite{BogKeat}.

The interesting consequence of the above formula is the expression for the
2-point form-factor  
\begin{equation}
K^{(off)}(t)=\frac{1}{4\pi^2}
\sum_{(p,q)=1}\left (\frac{\mu(q)}{\psi(q)}\right )^2(\frac{q}{p})
\delta (t- \bar{d}-\frac{1}{2\pi}\ln\frac{q}{p}).
\end{equation}
This formula means that the off-diagonal 2-point form factor is a sum over
$\delta$-functions in  special points which are situated in a
vicinity of the Heisenberg time plus a difference of periods of two
pseudo-orbits (= logarithm of the difference between two integers). This set 
of $\delta$-functions is dense but the largest peaks correspond to the
shortest pseudo-orbits. Similarly the 2-point diagonal form factor is the
sum of $\delta$ functions in the positions of periodic orbits
\begin{equation}
K^{(diag)}(t)=\frac{1}{4\pi^2}\sum_{p,m}\frac{\ln^2 p}{p^m}\delta
(t-\frac{m}{2\pi}\ln p).
\end{equation}
The smooth values corresponding to the random matrix predictions appears only
after a smoothing of these functions over a suitable interval of $t$.

\subsection{\bf Arithmetical systems}\label{arithmetical}

Similar behavior has been observed in a completely different model, namely for
distribution of eigenvalues of the Laplace--Beltrami operator for the
modular domain \cite{Arithmetic}. Using a generalization of the
Hardy-Littlewood method  it was shown that in this model the 2-point
correlation form factor can be written in the following form
\begin{equation}
K(t)=\frac{1}{\pi^3 k}\sum_{(p,q)=1}\left |\frac{q}{p}\beta (p,q)\right|^2
\delta (t-t_{p,q}),
\end{equation}
where 
\begin{equation}
t_{p,q}=\frac{2}{k}\ln \frac{kq}{\pi p},
\end{equation}
and 
\begin{equation}
\beta(p,q)=\frac{S(p,p;q)}{q^2\prod_{\omega|q}(1-\omega^{-2})}.
\end{equation}
Here the product is taken over all prime divisors of $q$ and $S(p,p;q)$ is
the Kloosterman sums
\begin{equation}
S(n,m;c)=\sum_{(d,c)=1}\exp \left (\frac{2\pi}{c}(nd+md^{-1})\right ).
\end{equation}
This model belongs to the so-called arithmetical models which are models on
the constant curvature surfaces generated by discrete arithmetic groups. For
all these models due to the exponential multiplicity of periodic orbits  one 
expects \cite{Georgeot}
that the spectral statistics will tend to the Poisson distribution though
from classical point of view all these models are the best known examples of
classically chaotic motion. Using the above expression one can prove this
statement for the modular domain.

\section{Construction of the density of states from finite number of periodic
  orbits}\label{finite}

The main difficulty in using the trace formulas is their divergent character. 
They cannot converge on the real axis as they have
to produce $\delta$-functions singularities there. Usually they are defined
by an analytical continuation from a region in the complex plane of energy
which do not touch the real axis. E.g. the Riemann zeta function converges
only when Re $ s>1$ but zeros are assumed to lie on the axis Re $s=1/2$.
The diagonal approximation consists in some sense on the computing the density
of states from a sum over a finite number of periodic orbits but such sums
can never have  $\delta$-function singularities. In this section we shall
discuss a special method \cite{Beyond}
which permits to avoid this difficulty and  produces an
(artificial) expression for the density of states with required
singularities from the knowledge of finite number of periodic orbits.

Let us write the semi-classical formula not for the density of states but for
the staircase function
\begin{equation}
N_{T^*}(E)=\bar{N}(E)+N^{(osc)}_{T^*}(E),
\end{equation}
where the oscillating part of this function is truncated (smoothly if
necessary) so to include periodic orbits with period up  to a fixed
period $T^*$, where $T^*$ is a parameter to be fixed later
\begin{equation}
N^{(osc)}_{T^*}(E)=2\sum_{T_p<T^*}\sum_{n=1}^{\infty}\tilde{A}_{p,n}
\sin(n(\frac{S_p}{\hbar}-\frac{\pi}{2}\mu_p)),
\end{equation}
and  for dynamical systems  the pre-factor is 
\begin{equation}
\tilde{A}_{p,n}=\frac{1}{2\pi |\det (M_p^n-1)|^{1/2}},
\end{equation}
and for the Riemann zeta function
\begin{equation}
\tilde{A}_{p,n}=-\frac{1}{2\pi n p^{n/2}},
\end{equation}
In the limit $T^*\rightarrow \infty$  function $N_{T^*}(E)$ should have a unit
jump each time when  $E$ equals an eigenvalue. For finite value of $T^*$
one can at the best have a smooth increase at these points. The idea
of the proposed method  is to define  semi-classical
eigenvalues  $E_n$ according to the following `quantization condition' \cite{Steiner}
\begin{equation}
N_{T^*}(E_n)=n+\frac{1}{2}.
\label{NT}
\end{equation}
The main advantage of this method is that it cannot miss any one
level simply because lines $N=n+1/2$ will  in any case cross the curve $N(E)$
and will produce (approximate) semi-classical energy levels. In
principle, one can obtain additional levels if the curve $N_{T^*}(E)$ has 
decreasing parts. But for not too big number of included orbits 
it is not the case and  this method  is quite efficient for numerical
computations of semi-classical levels \cite{BogSchmit}.

The other important point (which explain why we take into account the
infinite sum over repetitions of periodic orbit) is  that
\begin{equation}
\exp (2\pi i N_{T^*}(E))=e^{2\pi i \bar{N}}\frac{z_{T^*}^*(E)}{z_{T^*}(E)},
\end{equation}
where
\begin{equation}
z_{T^*}(E)=\prod_{T_p<T^*}\prod_{m=0}^{\infty}
(1-\frac{e^{iS_p/\hbar-i\pi\mu_p/2}}{|\Lambda_p |^{1/2}\Lambda_p^m}),
\end{equation}
is a truncated product over periodic orbits and the quantization condition
(\ref{NT}) is equivalent  to calculation of
semi-classical energy levels $E_n$ from the condition
\begin{equation}
Z_{T^*}(E_n)=0,
\label{zeta}
\end{equation}
where $Z_{T^*}(E)$ is a special form of the dynamical zeta function
\begin{equation}
Z_{T^*}(E)=z_{T^*}(E)+e^{2\pi i \bar{N}}z_{T^*}^*(E).
\end{equation}
The above expression has a Riemann--Siegel form \cite{BerryKeating} and 
automatically obeys the important functional equation for the zeta function 
\begin{equation}
Z(E)=e^{2\pi i \bar{N}}Z^*(E),
\end{equation}
which is crucial in proving the correct  analytical properties of the zeta
function.

Let us define a new bootstrapped density 
\begin{equation}
D_{T^*}(E)=\sum_n\delta(E-E_n),
\end{equation}
where instead of exact energy levels we use the semi-classical ones defined
as solutions of quantization condition (\ref{NT}) (or (\ref{zeta}). Rewriting 
$D_{T^*}(E)$ in the form
\begin{equation}
D_{T^*}(E)=d_{T^*}(E)\sum_n \delta (N_{T^*}(E)-n-1/2),
\end{equation}
where $d_{T^*}(E)=d N_{T^*}(E)/dE$
and using the Poisson summation formula one gets
\begin{equation}
D_{T^*}(E)=d_{T^*}(E)\sum_{k=-\infty}^{\infty} (-1)^k
\exp (2\pi ik N_{T^*}(E)).
\end{equation}
We called it a bootstrapped density as it  has all Fourier harmonics
and not only the ones with period $T<T^*$, the role of the effective orbits
introduced beyond $T^*$ being to generate the correct analytical properties  
associated with the discreteness of quantum spectrum. 

Substituting this expression to the formula for the 2-point correlation
function one gets
\begin{eqnarray}
R_2(\epsilon_1,\epsilon_2)&=&< d_{T^*}(E+\epsilon_1)d_{T^*}(E+\epsilon_2)
\sum_{k_1,k_2}(-1)^{k_1-k_2}
\nonumber\\
&& \times \exp (2\pi i (k_1 N_{T^*}(E+\epsilon_1)-N_{T^*}(E+\epsilon_2)))>.
\end{eqnarray}
Consider first the $k_1=k_2=0$ term, which we write in the form
\begin{equation}
< d_{T^*}(E+\epsilon_1)d_{T^*}(E+\epsilon_2)>=\bar{d}^2+
R_2^{(diag)}(\epsilon_1,\epsilon_2),
\end{equation}
where
\begin{equation}
R_2^{(diag)}(\epsilon_1,\epsilon_2)=\frac{1}{4\pi^2}
\frac{\partial^2}{\partial \epsilon_1 \partial \epsilon_2}
\ln \Delta (\epsilon_1,\epsilon_2),
\end{equation}
and
\begin{equation}
\ln \Delta (\epsilon_1,\epsilon_2)=4\pi^2
<N_{T^*}^{(osc)}(E+\epsilon_1)N_{T^*}^{(osc)}(E+\epsilon_2))>.
\end{equation}
Because only a finite number of orbits enters this expression we assume 
that the resulting sum over pairs of periodic orbits can be replaced by the
diagonal terms, for which only orbits with exactly the same actions will
contribute. The result can be expressed in the form
\begin{equation}
\Delta (\epsilon_1,\epsilon_2)=|Z_g(i\epsilon)|^2,
\end{equation}
where
\begin{equation}
Z_g(s)=\prod_{T_p<T^*} \prod_{n=1}^{\infty}
\exp( \frac{4\pi^2}{n^2}g_p|A_{p,n}|^2e^{nT_p s}),
\end{equation}
and $g_p$ is the number of orbits with period $T_p$ (multiplicity of
periodic orbit).

In generic systems the multiplicity $g_p$ is the same for almost all orbits
and so, if its value is denoted $g$
\begin{equation}
Z_g(s)=Z^g(s),
\end{equation}
where $Z(s)$ is defined by the same formula but with $g_p=1$. For systems
without time-reversal invariance $g=1$, and for systems whose dynamics is
time-reversal symmetric $g=2$.

Since the $k_1=k_2=0$ term corresponds to the usual diagonal approximation,
the other terms are representing the off-diagonal contributions, $R_2^{(off)}$. 
To evaluate them we note that the Taylor expansion in powers of $\epsilon$ of 
the mean staircase function leads to a
term $(k_1-k_2)\bar{N}(E)$ in the phase which is of the order of $\hbar^{-f}$.
Hence the energy average renders negligible any contributions with 
$k_1\neq k_2$, and therefore
\begin{equation}
R_2^{(off)}(\epsilon)= \frac{\partial^2}
{\partial \epsilon_1 \partial \epsilon_2}
\sum_{k\neq 0}\frac{1}{(2\pi k)^2}
\exp (2\pi i \bar{d} k (\epsilon_1-\epsilon_2)) \Phi_k(\epsilon_1, \epsilon_2),
\end{equation}
where
\begin{equation}
\Phi_k(\epsilon_1,\epsilon_2)=<\exp (2\pi i k 
(N_{T^*}^{(osc)}(E+\epsilon_1)-N_{T^*}^{(osc)}(E+\epsilon_2)))>.
\end{equation}
We now make the key assumption that in generic systems the orbits up to
period $T^*$ are non-commensurable modulo exact
degeneracies, and hence that the energy average of any smooth function of
$\exp (iS_p)$
\begin{equation}
<f>=<f(e^{iS_1(E)},e^{iS_2(E)},\ldots,e^{iS_M(E)})>,
\end{equation}
can be calculated using
\begin{equation}
<f>=\int_0^{2\pi}\ldots \int_0^{2\pi}
f(e^{i\phi_1},\ldots,e^{i\phi_M})\prod_j^M\frac{d\phi_j}{2\pi}.
\end{equation}
This is essentially equivalent to  random phase approximation, or to
ergodic theorem for quasi-periodic functions with non-commensurate periods,
or to strict diagonal approximation. 

Using the trace formula for the staircase function one can thus perform the
energy average by evaluating these integrals. Exact results for certain cases
will be discussed later. First, for clarity, we consider a simple leading
order approximation to $\Phi_k$ based on the relation
\begin{equation}
<\exp (i G(E))>\approx \exp (-\frac{1}{2}<G^2(E)>).
\end{equation}
This is an identity if $G$ is a Gaussian random function with zero mean. But
if one ignores all terms with repetitions of the same periodic orbits the
formula for $N^{(osc)}(E)$ will be equal to a sum of a big number of terms
with non-commensurable frequencies which can be considered 
as independent random variables therefore  the resulting distribution of 
$N^{(osc)}(E)$ being a sum over many independent random variables should a Gaussian
distributed random function. Consequently this type of approximation is 
expected to  be a good approximation for generic systems. 

From it one obtains
\begin{eqnarray}
\Phi_k(\epsilon_1,\epsilon_2)&=&<\exp (-2\pi^2  k^2
<(N_{T^*}^{(osc)}(E+\epsilon_1)-N_{T^*}^{(osc)}(E+\epsilon_2))^2)>
\nonumber\\
&=& (\frac{\Delta(\epsilon)}{L^2})^{k^2},
\end{eqnarray}
where $\Delta (\epsilon)$ was defined above and 
\begin{equation}
L=\exp (2\pi^2<(N^{(osc)}(E))^2>).
\end{equation}
Using the Hannay-Ozorio de Almeida sum rule \cite{HannayOzorio}, it may 
be shown that
\begin{equation}
L\approx (T^*)^g.
\end{equation}
Since we anticipate taking $T^*\approx T_H$, it then follows that the terms
in the $k$ sum decrease rapidly as $(\bar{d})^{-2gk^2}$. Hence in the
leading semi-classical order as $\epsilon \bar{d} \rightarrow \infty$, we may
retain just the $k=\pm 1$ contributions and when deriving over $\epsilon$
take into account only terms with $\exp (2\pi i \bar{d}\epsilon)$. Finally 
one gets
\begin{equation}
R_2^{(off)}(\epsilon_1,\epsilon_2)=\bar{d}^2e^{2\pi i \bar{d}\epsilon}
<\frac{z_{T^*}^*(E+\epsilon_1)z_{T^*}(E+\epsilon_2)}
{z_{T^*}(E+\epsilon_1)z_{T^*}^*(E+\epsilon_2)}> + c.c.
\end{equation}
The last step consists in performing the exact average in this formula. We
shall consider the case of the Riemann zeta function first.

\subsection{Off-diagonal terms for the Riemann zeta
  function}\label{offriemann}

For the Riemann zeta function the  the ratio of four
truncated zeta functions in the above formula has the form
\begin{equation}
<\frac{z_{T^*}^*(E+\epsilon_1)z_{T^*}(E+\epsilon_2)}
{z_{T^*}(E+\epsilon_1)z_{T^*}^*(E+\epsilon_2)}>=<\prod_{\ln p<T^*}R_p>,
\label{126}
\end{equation}
where
\begin{equation}
R_p(\phi_p)=\frac{(1-A_p e^{i\phi_p+i\tau_p^{(1)}})(1-A_p e^{-i\phi_p-i\tau_p^{(2)}})}
{(1-A_p e^{-i\phi_p-i\tau_p^{(1)}})(1-A_p e^{i\phi_p+i\tau_p^{(2)}})},
\end{equation}
and[B
\begin{equation}
A_p=\frac{1}{\sqrt{p}},\;\;\phi_p=E\ln p,\;\;\tau_p^{(i)}=\epsilon_i \ln p. 
\end{equation}
As the logarithms of prime numbers are non-commensurable,  $e^{i\phi_p}$
act as independent random variables for each $p$ and 
\begin{equation}
<\prod_{\ln p<T^*}R_p>=\prod_{\ln p<T^*}<R_p>.
\end{equation}
The mean value of an individual $R_p$ is
\begin{eqnarray}
<R_p>&=&\int_0^{2\pi}R_p(\phi )\frac{d\phi_p}{2\pi}
\nonumber\\
&=&\frac{1}{2\pi i}\oint \frac{(1-A_p z e^{i\tau_p^{(1)}})
  (1-A_pz^{-1}e^{-i\tau_p^{(2)}})}
{(1-A_p z^{-1} e^{-i\tau_p^{(1)}})(1-A_p z e^{i\tau_p^{(2)}})}
\frac{dz}{z},
\end{eqnarray}
where the integral is taken over a unit circle in the complex $z$-plane.

The last integral is easily computed by the residues. There is two poles
inside the contour. The first is at $z=0$ and the second one is at 
$z=A_pe^{-i\tau_p^{(1)}}$. Simple algebra gives
\begin{equation}
<R_p>=\frac{(1-A_p^2)^2}{|1-A_p^2e^{i\tau_p}|^2}\left (1-
\frac{A_p^4(1-e^{i\tau_p})^2}{(1-A_p^2)^2}\right ),
\end{equation}
where $\tau_p=\tau_p^{(1)}-\tau_p^{(2)}$.

The total contribution equals the product over all primes up to $T^*$
\begin{equation}
R_2(\epsilon) =C^2\exp (2\pi i \bar{d}\epsilon) 
|\zeta(1+i\epsilon)|^2 \Phi^{(off)}(\epsilon),
\label{132}
\end{equation}
where
\begin{equation}
\zeta (s)=\prod_p\frac{1}{1-p^{-s}},\; 
\Phi^{(off)}(\epsilon)=\prod_p(1-\frac{(1-p^{i\epsilon})^2}{(p-1)^2}),
\end{equation}
and
\begin{equation}
C=\bar{d}\prod_p(1-\frac{1}{p}).
\end{equation}
All products in these expressions include prime numbers up to $\ln p=T^*$.
The two first products converge when $T^*\rightarrow \infty$ and only the
last one require regularisation. But our parameter $T^*$ has not yet been
fixed. Let us choose it in such a way that
\begin{equation}
C\equiv \bar{d}\prod_{\ln p <T*}(1-\frac{1}{p})=\frac{1}{2\pi}.
\label{135}
\end{equation}
It is easy to check \cite{Titchmarsh} that asymptotically this equation leads
\begin{equation}
T^*=2\pi \bar{d} e^{\gamma},
\end{equation}
where $\gamma$ is the Euler constant. Therefore, our $T^*$ is of the order
of the Heisenberg time as was expected. Note that exactly the same factor
$e^{\gamma}$ often appears in the statistical approach to prime numbers
(see discussion in \cite{HardyLittlewood}) and can be considered as a
renormalisation of formally divergent sums.

After this renormalisation we get exactly the same formula as has been
derived in the previous section using the Hardy-Littlewood conjecture about
the pairwise distribution of prime numbers. Note that in a present derivation 
no analogous conjectures have been assumed.

\subsection{Off-diagonal contribution for dynamical
  systems}\label{offdynamical}

Let us now compute off-diagonal contribution to dynamical systems. Our
starting point will be the same expression as for the Riemann zeta function
\begin{equation}
R_2^{(off)}(\epsilon_1,\epsilon_2)=\bar{d}^2e^{2\pi i \bar{d}\epsilon}
<\frac{z_{T^*}^*(E+\epsilon_1)z_{T^*}(E+\epsilon_2)}
{z_{T^*}(E+\epsilon_1)z_{T^*}^*(E+\epsilon_2)}>.
\end{equation}
The only difference with Riemann case is that the truncated zeta function
$z_{T^*}(E)$ contains now an infinite product over $m$
\begin{equation}
z_{T^*}(E)=\prod_{T_p<T^*}\prod_{m=1}^{\infty}
(1-\frac{e^{iS_p/h-i\pi/2\mu_p}}
{|\Lambda_p|^{1/2}\Lambda_p^{m}}).
\label{142}
\end{equation}
As above we shall assume that all periods of primitive periodic orbits are
non-commen\-su\-rable. Therefore $e^{iT_p\epsilon}$ can be considered as
independent random variables but one cannot ignore the existence of
products over $m$ in Eq.~(\ref{142}) 
\begin{equation}
<\frac{z_{T^*}^*(E+\epsilon_1)z_{T^*}(E+\epsilon_2)}
{z_{T^*}(E+\epsilon_1)z_{T^*}^*(E+\epsilon_2)}>=\prod_{T_p<T^*}<R_p(\phi_p)>,
\end{equation}
where
\begin{equation}
<R_p>=\int_0^{2\pi}R_p(\phi_p)\frac{d\phi_p}{2\pi},
\end{equation}
and
\begin{equation}
R_p(\phi_p)=\prod_{m=1}^{\infty}
\frac{(1-A_{p,m} e^{-i\phi_p-i\tau_p^{(1)}})
(1-A_{p,m} e^{i\phi_p+i\tau_p^{(2)}})}
{(1-A_{p,m} e^{i\phi_p+i\tau_p^{(1)}})(1-A_{p,m} e^{-i\phi_p-i\tau_p^{(2)}})},
\end{equation}
and
\begin{equation}
A_{p,m}=|\Lambda_p|^{-1/2}\Lambda_p^{-m},\;\phi_p=S_p(E)/\hbar-\pi \mu_p/2,
\;\tau_p^{(i)}=T_p \epsilon_i.
\end{equation}
 
Though it is possible to calculate the  integral $<R_p>$ by the residues as has 
been done in the previous Section it is more convenient to use the so-called 
$q$-binomial theorem (\cite{Gasper} p.7)
\begin{equation}
\sum_{n=0}^{\infty} \frac{(a;q)_n}{(q;q)_n}z^n=
\frac{(az;q)_{\infty}}{(z;q)_{\infty}}.
\end{equation}
Here
\begin{equation}
(a;q)_n=(1-a)(1-aq)(1-aq^2)\ldots (1-aq^{n-1}),
\end{equation}
and
\begin{equation}
(a;q)_{\infty}=\prod_{n=0}^{\infty}(1-aq^n).
\end{equation}
Denoting 
\begin{equation}
q=\Lambda_p^{-1},\; z=|\Lambda_p|^{-1/2}e^{-i\phi_p-i\tau_p^{(2)}},
\;a=e^{i\tau_p^{(2)}-i\tau_p^{(1)}},
\end{equation}
one obtains
\begin{equation}
\prod_{m=0}^{\infty}\frac{(1-A_{p,m}e^{-i\phi_p-i\tau_p^{(1)}})}
     {(1-A_{p,m}e^{-i\phi_p-i\tau_p^{(2)}})}=
\frac{(az;q)_{\infty}}{(z;q)_{\infty}}.
\end{equation}
Similarly
\begin{equation}
\prod_{m=0}^{\infty}\frac{(1-A_{p,m}e^{i\phi_p+i\tau_p^{(2)}})}
     {(1-A_{p,m}e^{i\phi_p+i\tau_p^{(1)}})}=
\frac{(az';q)_{\infty}}{(z';q)_{\infty}},
\end{equation}
where $z'=|\Lambda_p|^{-1/2}e^{i \phi_p+i\tau_p^{(2)}}$.

Using the $q$-binomial theorem these expressions can be represented as follows
\begin{equation}
\prod_{m=0}^{\infty}\frac{(1-A_{p,m}e^{-i\phi_p-i\tau_p^{(1)}})}
     {(1-A_{p,m}e^{-i\phi_p-i\tau_p^{(2)}})}=
\sum_{n=0}^{\infty}\frac{(a;q)_{n}}{(q;q)_{n}}z^n,
\end{equation}  
and
\begin{equation}
\prod_{m=0}^{\infty}\frac{(1-A_{p,m}e^{i\phi_p+i\tau_p^{(2)}})}
     {(1-A_{p,m}e^{i\phi_p+i\tau_p^{(1)}})}=
\sum_{n=0}^{\infty}\frac{(a;q)_{n}}{(q;q)_{n}}z'^n.
\end{equation}
Now  $<R_p>$ is the integral of the product of these two expressions and in
the diagonal approximation only terms with the same power will contribute and 
finally 
\begin{equation}
<R_p>=\sum_{n=0}^{\infty}\frac{(a;q)_n^2}{(q,q)_n^2} y^n,
\end{equation}
where $y=|\Lambda_p|^{-1}e^{i(\tau_p^{(1)}-\tau_p^{(2)})}$. This sum may be
expressed through the so-called $q$-hypergeometric function \cite{Gasper}
\begin{equation}
<R_p>
=_2\phi_1(e^{-iT_p\epsilon },e^{-iT_p\epsilon};\Lambda_p^{-1};\Lambda_p^{-1},
|\Lambda_p|^{-1}e^{iT_p\epsilon}),
\end{equation}
where we take into account that $\tau_p^{(1)}-\tau_p^{(2)}=T_p(\epsilon)$.

The total mean value equals the product over all periodic orbits till $T^*$.
The divergent part comes only from  $n=1$ term  
\begin{equation}
\prod_{T_p<T^*}(1+\frac{1}{|\Lambda_p|}(e^{iT_p\epsilon}+e^{-iT_p\epsilon}-2)).
\end{equation}
But the divergence of this product coincides with the divergent part of
the expression
\begin{equation}
\prod_{T_p<T^*}\left |\frac{Z_p(0)}{Z_p(i\epsilon)}\right |^2,
\end{equation}
where
\begin{equation}
Z_p(s)=1-\frac{e^{T_ps}}{|\Lambda_p|}.
\end{equation}
The product
\begin{equation}
\prod_{T_p<T^*}<R_p>\left |\frac{Z_p(i\epsilon)}{Z_p(0)}\right |^2  
\end{equation}
converges when $T^*\rightarrow \infty$ and
\begin{equation}
R_2(\epsilon)= \frac{e^{2\pi i\bar{d}\epsilon}}{4\pi^2}
  |\gamma^{-1}Z_{cl}(i\epsilon)|^2 \Phi^{(osc)}(\epsilon) +c.c.,
\label{157}  
\end{equation}
where
\begin{equation}
\Phi^{(osc)}(\epsilon)=\prod_{p}<R_p>
\left |\frac{Z_p(i\epsilon)}{Z_p(0)}\right |^2.  
\label{151}
\end{equation}
and $Z_{cl}(s)$ is a classical zeta function (see the footnote at page
\pageref{footnote}) defined when Re$s<0$ by the product over all ppo
\begin{equation}
Z_{cl}(s)=\prod_{ppo}(1-\frac{e^{T_ps}}{|\Lambda_p|})^{-1}.
\label{159}
\end{equation}
Let us fix the maximal period $T^*$ from the condition
\begin{equation}
\bar{d}\prod_{T_p<T^*}Z_p(0)=\frac{1}{2\pi |\gamma|},
\end{equation}
where $\gamma$ is the residue of $Z_{cl}(s)$ at $s=1$ (which existence is a
consequence of Hannay-Ozorio de Almeida sum rule) 
\begin{equation}
\gamma =\lim_{s\rightarrow 0}sZ_{cl}(s).
\end{equation}
As above this renormalisation defines  $T^*$ of the order of $T_H$ and ensures
that when $\epsilon \rightarrow 0$ $R_2(\epsilon)$ tends to the oscillatory
part of the GUE result (\ref{25b})
\begin{equation}
R_2^{(off)}(\epsilon)=\frac{1}{4\pi^2\epsilon^2}(e^{2\pi i \bar{d}\epsilon}+
e^{-2\pi i \bar{d}\epsilon}).
\end{equation}
In principle any zeta function whose divergent part at real $\epsilon$
coincides with $Z_{cl}(i\epsilon)$ can be used in the above expressions.

\section{Random matrix universality}\label{universal}

This section is devoted to another method of semi-classical calculation of 
off-diagonal part of correlation function based on a different idea.

In is well known \cite{Mehta} that the standard ensembles of random matrices
correspond to the following choice of measure in the ensemble
\begin{equation}
P(M)=\exp (-\mbox{Tr}V(M)),
\end{equation}
where $V(M)$ is an arbitrary function. The important universality in the
random matrices ensembles is the fact that the unfolded distribution does not
depend on the explicit form of this function \cite{Brezin} provided that it 
corresponds to the so-called definite momentum problem \cite{Pato}.

For unitary ensembles all correlation functions have  the form \cite{Mehta}
\begin{equation}
R_n(x_1,\ldots,x_n)=\det |K_N(x_i,x_j)|_{i,j=1,\ldots, n},
\end{equation}
where the kernel
\begin{equation}
K_N(x,y)=e^{-V(x)/2-V(y)/2}\sum_{n=1}^{N}p_n(x)p_n(y),
\end{equation}
and $p_n(x)$ are polynomials of degree $n-1$ orthogonal with respect to the
measure $\exp (-V(x))$
\begin{equation}
\int p_n(x)p_m(x)e^{-V(x)}dx=\delta_{nm}.
\end{equation}
The `semi-classical' asymptotic of orthogonal polynomials leads to the
following expression for the kernel $K(x,y)=K_N(x,y)$ when $N\rightarrow \infty$
\begin{equation}
K(x,y)=\frac{\sin \pi(N(x)-N(y))}{\pi (x-y)},
\end{equation}
where $N(x)=\int^{x} \rho(x')dx'$ is a mean staircase function related 
to  $V(x)$ by the Dyson equation
\begin{equation}
\int \frac{\rho (t)}{x-t}dt=\frac{1}{2}V'(x).
\label{168}
\end{equation}
The mean density $\rho(x)$ does depend on the form of $V(x)$ but if
\begin{center}
$\epsilon=x-y\ll$ characteristic scale of the potential $V(x)$
\end{center}
one can expand the
difference $N(x)-N(y)$ into powers of $\epsilon$, $N(x)-N(y)=\bar{d}\epsilon$,
and all correlation functions will be functions of $\bar{d}\epsilon$ which
ensures their universality after rescaling. 

Now we shall use this fact in addition to the trace formula.

Let us assume that we know all periodic orbits up to   period $T^*$ but,
contrary to what was used in the previous sections, this maximal period will
be much smaller than the Heisenberg time but still much bigger than pure  
classical time
\begin{equation}
t_{cl}\ll T^*\ll t_H.
\label{169}
\end{equation}
Below we assume that the maximal period is of the order of the
Ehrenfest time 
\begin{equation}
T^*\approx t_E.
\end{equation}
According to the trace formula the density of states has the form
\begin{equation}
d(E)=\tilde{d}(E)+\eta(E),
\end{equation}
where as above
\begin{equation}
\tilde{d}(E)=  \bar{d}+\sum_{T_p<T^*}\sum_{n=1}^{\infty}
\frac{T_p}{\hbar}(A_{p,n}e^{i n (S_p/\hbar -\pi\mu_p/2)}+c.c.),
\end{equation}
and $\eta(E)$ is (unknown) part of the density constructed form high-period
orbits. 

Let us now try  to construct a random matrix ensemble which has the
mean density  exactly equals $\tilde{d}(E)$. In principle the necessary
potential can be computed from the Dyson equation (\ref{168}). But the explicit 
form of this potential is irrelevant as all correlation functions depend only 
on the mean staircase function $\tilde{N}(x)=\int^x \tilde{d}(x')dx'$ which is 
supposed to be known.

Hence, the 2-point correlation function in such ensemble  takes the form
\begin{equation}
\tilde{R}_2(x,y)=\tilde{d}(E+x)\tilde{d}(E+y)-
\frac{\sin^2(\pi (\tilde{N}(E+x)-\tilde{N}(E+y)))}{(\pi(x-y))^2}.
\end{equation}
$\tilde{d}(E)$ in this formula is the mean (over constructed ensemble)
density of eigenvalues  and it is not a constant but includes periodic orbits
with period ut to $T^{*}$ and consequently it has oscillations in the scale
smaller or equal $T^{*}$. To be consistent with the standard definition of
the 2-point correlation function (\ref{34}) it is necessary to perform an
additional smoothing over over an energy window, $\Delta E$,   much bigger
than $\hbar/T^*$
\begin{equation}
\frac{T^*}{\hbar}\Delta E \gg 1,
\end{equation}
Denoting this average by the brackets $<\ldots>$, the 2-point
correlation function  
\begin{equation}
R_2(x,y)=<\tilde{d}(E+x)\tilde{d}(E+y)-
\frac{\sin^2(\pi \tilde{N}(E+x)-\pi \tilde{N}(E+y))}{\pi^2(x-y)^2}>.
\end{equation}
Separating the smooth and oscillatory parts one gets
\begin{eqnarray}
R_2^{(diag)}(x,y)&=&<\tilde{d}(E+x)\tilde{d}(E+y)>-\frac{1}{2\pi^2(x-y)^2}
\nonumber\\
&=&\bar{d}^2+\frac{1}{4\pi^2}
\frac{\partial^2}{\partial x\partial y}\ln \Delta (x,y),
\end{eqnarray}
where 
\begin{equation} 
  \Delta(x,y)=\frac{1}{(x-y)^2}\exp(-4\pi^2<\tilde{N}(E+x)\tilde{N}(E+y)>),
\end{equation}
and
\begin{equation}
R_2^{(off)}(x,y)=\frac{1}{4\pi^2(x-y)^2}
<e^{2\pi i (\tilde{N}(E+x)-\tilde{N}(E+y))}>+c.c.
\end{equation}
As  above we assume that all periodic orbit actions in the explicit 
expression for $\tilde{d}(E)$ are non-commensurable. Because we chose $T^*$ 
to be of the order of the Ehrenfest time, one can perform the smoothing over
energy window in the strict diagonal approximation.
But all the expressions which have to smooth are exactly the same as were above.
The only difference is that the limiting period $T^*$ is of the order not of the
Heisenberg time but of the Ehrenfest time and there is a factor $(x-y)^{-2}$. 
Using the same transformations as above one concludes that
\begin{equation}
  \Delta(\epsilon)=\frac{1}{\epsilon^2}|Z_{cl}(i\epsilon,T^*)|^2
  \Phi^{(diag)}(\epsilon),
\end{equation}  
and
\begin{equation}
   R_2^{(off)}(\epsilon)=\frac{1}{4\pi^2\epsilon^2}
   \left |\frac{Z_{cl}(i\epsilon,T^*)}{Z_{cl}(0,T^*)}\right |^2
  \Phi^{(off)}(\epsilon).
\label{182}
\end{equation}
Here $\epsilon=x-y$ and functions $\Phi^{(diag)}(\epsilon)$ and
$\Phi^{(off)}(\epsilon)$ are defined exactly as Eqs.~(\ref{40}) and (\ref{151})
except that they are truncated at $T^*$. Because these products converge 
and the value of  $T^*$ is assumed to obeys inequalities (\ref{169}) one
can go to the limit $T^*\rightarrow \infty$ and the resulting expressions
will coincide with Eqs.~(\ref{40}) and (\ref{151}).

The function $Z_{cl}(s,T^*)$ is the truncated product 
\begin{equation}
Z_{cl}(s,T^*)=\prod_{T_p<T^*}(1-\frac{e^{sT_p}}{|\Lambda_p|})^{-1}.
\end{equation}
Because $T^*$ is such that this product includes a large number of terms
one can consider the behavior of the product in the limit $T^*\rightarrow \infty$.

Let us perform the following formal transformations (which can be done
rigorously by adding a small imaginary part to $\epsilon$
\begin{equation}  
   \frac{Z_{cl}(i\epsilon,T^*)}{\epsilon Z_{cl}(0,T^*)}
   \equiv \frac{1}{\epsilon}\prod_{T_p<T^*}\frac{1-\frac{1}{|\Lambda_p|}}
   {1-\frac{e^{i\epsilon T_p}}{|\Lambda_p|}}=\frac{1}{\gamma}Z_{cl}(i\epsilon),
\end{equation}
where $Z_{cl}(i\epsilon)$ is the classical zeta function (\ref{159}) defined
as the product over ppo with arbitrary period and 
\begin{equation}
   \gamma=\epsilon\prod_{T_p<T^*}\frac{1}{1-\frac{1}{|\Lambda_p|}}
   \prod_{T_p>T^*}\frac{1}{1-\frac{e^{i\epsilon T_p}}{|\Lambda_p|}}.
\end{equation}
But 
\begin{eqnarray}
   && \prod_{T_p<T^*}\frac{1}{1-\frac{1}{|\Lambda_p|}}=\lim_{s\rightarrow 0}
   \prod_{T_p<T^*}\frac{1}{1-\frac{e^{isT_p}}{|\Lambda_p|}}
\nonumber\\
   &&=\lim_{s\rightarrow 0}Z_{cl}(is)
   (\prod_{T_p>T^*}(1-\frac{e^{isT_p}}{|\Lambda_p|}).
\end{eqnarray}
Therefore
\begin{equation}
   \gamma=\epsilon\lim_{s\rightarrow 0}Z_{cl}(is)
   \prod_{T_p>T^*}\frac{(1-\frac{e^{isT_p}}{|\Lambda_p|})}
   {(1-\frac{e^{i\epsilon T_p}}{|\Lambda_p|})}.
\end{equation}
When $T^*\rightarrow \infty$
\begin{equation}
   \prod_{T_p>T^*}(1-\frac{e^{isT_p}}{|\Lambda_p|}) \rightarrow
   \exp (-\sum_{T_p>T^*}\frac{e^{isT_p}}{|\Lambda_p|}).
\end{equation}
Using Hannay-Ozorio de Almeida sum rule \cite{HannayOzorio} one
concludes that  when $T^*\rightarrow \infty$ this expression is close to
\begin{equation}
   \exp (-\int_{T^*}^{\infty}\frac{e^{isT}}{T}dT).
\end{equation}
Finally for large $T^*$ 
\begin{equation}
   \gamma=\epsilon\lim_{s\rightarrow 0}Z_{cl}(is)
   \exp (-\int_{T^*}^{\infty}\frac{e^{i\epsilon T}}{T}dT+
   \int_{T^*}^{\infty}\frac{e^{i s T}}{T}dT).   
\end{equation}
As the limiting period  has been chosen such that $\epsilon T^* \ll 1$ (and
of course $sT^* \ll 1$) one can use the approximation
\begin{equation}
\int_{T^*}^{\infty}\frac{e^{i\epsilon T}}{T}dT\rightarrow \ln
(T^*\epsilon)+\mbox{const},
\end{equation}
Therefore
\begin{equation}
   \gamma=\lim_{s\rightarrow 0}sZ_{cl}(is),
\end{equation}
and as only $|\gamma |$ is important Eq.~(\ref{182}) coincides with 
Eq.~(\ref{157}) derived by completely different method.

\section{Riemann-Siegel form of  density of states}\label{RiemannSiegel}

In this Section we propose  one more method of computation spectral
statistics based on ideas different from the ones discussed in the previous
Sections. For simplicity we consider only the Riemann case.   

It is known that the best method of calculation of the Riemann zeta function
is the using the Riemann-Siegel representation of it \cite{Titchmarsh}, 
\cite{BerryKeating} (called in simple cases the approximative functional equation).

Heuristically it can be derived  as follows. 
Let us divide the sum over integers in the definition of the Riemann zeta 
function into two contributions
\begin{equation}
\zeta(1/2-iE)=\sum_{n=1}^{\infty}\frac{1}{n^{1/2}}e^{iE\ln n}=
\sum_{n=1}^{N}\frac{1}{n^{1/2}}e^{iE\ln
  n}+\sum_{n=N+1}^{\infty}\frac{1}{n^{1/2}}e^{iE\ln n}.
\end{equation}
In the first sum  the summation is done over integers up to a (large) 
integer $N$ and the second sum  includes all terms with $n>N$. Using the
Poisson summation formula one can rewrite the latter in the form
\begin{equation}
\sum_{n=N+1}^{\infty}\frac{1}{n^{1/2}}e^{iE\ln n}=\sum_{m=-\infty}^{\infty}
\int_N^{\infty}\frac{1}{n^{1/2}}e^{-2\pi i m n+iE\ln n}dn,
\end{equation}
and it is naturally to apply the stationary phase method to compute the
integral.  The dominant contribution comes from  vicinities of points
$n_{sp}$ obeying the saddle-point equation
\begin{equation}
2\pi m  =\frac{E}{n_{sp}},
\end{equation}
or 
\begin{equation}
n_{sp}=\frac{E}{2\pi m},
\end{equation}
but this contribution exists only if this stationary point lies between $N$ and
$\infty$ which is equivalent to the following restriction on $m$
\begin{equation}
  0<m<N^*,
\end{equation}
where
\begin{equation}
N^*=\frac{E}{2\pi N}.
\end{equation}
Computing the second derivative of the phase and performing the integration
in the Gaussian approximation one gets
\begin{equation}
\sum_{n=N+1}^{\infty}\frac{1}{n^{1/2}}e^{iE\ln n}=e^{2\pi i\bar{N}(E)}
\sum_{n=1}^{N^*}\frac{1}{n^{1/2}}e^{-iE\ln n},
\end{equation}
where $\bar{N}(E)$ is  the mean staircase function given by Eq.~(\ref{17}).
(On more careful analysis see \cite{BerryKeating}).

The same result can be obtained without calculation (but less rigorously) by 
using the functional equation of the Riemann zeta function. It is known 
that this function obeys the relation \cite{Titchmarsh}
\begin{equation}
\zeta (1/2-iE)=e^{2\pi i \bar{N}(E)}\zeta (1/2+iE),
\end{equation}
which is a necessary condition that the complex valued function $\zeta(s)$
has zeros on a line. Substituting the formal expansion 
\begin{equation}
\zeta(1/2\pm iE)=\sum_{n=1}^{\infty}\frac{1}{n^{1/2}}e^{\pm i E\ln n}
\end{equation}
into this relation and comparing terms in two parts one concludes that
\begin{equation}
  \sum_{\ln n> T_1}\frac{1}{n^{1/2}}e^{ i E\ln n}=e^{2\pi i \bar{N}(E)}
  \sum_{\ln n<T_2} \frac{1}{n^{1/2}}e^{- i E\ln n},
\label{204}  
\end{equation}
provided $T_1+T_2=T_H$ where the Heisenberg time $T_H=2\pi \bar{d}(E)$.
One can also get a slightly more general relation
\begin{equation}
\sum_{n=N_1}^{N_2}\frac{1}{n^{1/2-iE}}
=e^{2\pi i\bar{N}(E)}\sum_{n=E/(2\pi N_2)}^{E/(2\pi N_1)}\frac{1}{n^{1/2+iE}}.
\end{equation}
This type of bootstrap where a sum of large integers is proportional
to a sum of the short ones is the cornerstone of the Riemann-Siegel method 
and it is very useful in calculation of zeta functions. Our purpose will be 
to use it to obtain an information about sums of primes.

Let us split the density of Riemann zeros (\ref{16}) into two contributions
\begin{equation}
d^{(osc)}(E)=d_1(E)+d_2(E).
\end{equation}
The first one includes the sum over all prime up to a certain integer $N$ 
\begin{equation}
d_1(E)=-\frac{1}{2\pi} \sum_{p<N}
\ln p\sum_{n=1}^{\infty}\frac{1}{p^{n/2}}e^{iEn\ln p} +c.c.
\end{equation}
and the second term contains all primes bigger than $N$
\begin{equation}
d_2(E)=-\frac{1}{2\pi} \sum_{p>N}
\ln p\sum_{n=1}^{\infty}\frac{1}{p^{n/2}}e^{iEn\ln p} +c.c.
\end{equation}
The main ingredient of following discussion is the sieve representation of
prime numbers. Namely we shall use the following relation
\begin{eqnarray}
  \sum_{p>N}f(p)&=&\sum_{n>N}f(n)-
  \sum_{p_j,n>N/p_j}f(p_j n)+
  \sum_{p_j,p_k;n>N/(pjp_k)}f(p_jp_k n)+\ldots \nonumber \\
&=& \sum_k \sum_{n>N/k}f(nk)\mu(k),  
\end{eqnarray}
where $\mu(k)$ is the Euler function.

This relation is just the manifestation of the fact that when one substitute
instead of a sum over prime numbers the sum over integers one has first to
subtract from this sum all numbers which are proportional to short primes but 
all numbers which are proportional to product of two primes are
subtracted twice, therefore, it is necessary to add integers proportional to the
product of two primes etc. This type of inclusion-exclusion principle is
exact provided one is restricted to a {\bf finite} collection of primes. 

Applying this formula to $f(n)=1/n^{1/2-iE}$ one gets
\begin{equation}
\sum_{p>N}\frac{1}{p^{1/2-iE}}=
\sum_{k=1}^{\infty}\sum_{n>N/k}\frac{1}{n^{1/2-iE}}
\frac{\mu(k)}{k^{1/2-iE}}.
\end{equation}
Using the approximate functional equation (\ref{204}) for the sums
over $n$ in this formula  one gets
\begin{equation}
\sum_{p=N_1}^{N_2}\frac{1}{p^{1/2-iE}}=e^{2\pi i\bar{N}(E)}
\sum_{k=1}^{\infty}\sum_{n=Ek/(2\pi N_2)}^{Ek/(2\pi N_1)}
\frac{1}{n^{1/2+iE}}\frac{\mu(k)}{k^{1/2-iE}}.
\end{equation}
Taking into account that the total `period' of an individual  term in the
double sum is $t=\ln k-\ln n$
this expression can be transformed  into the following form
\begin{equation}
\sum_{p=N_1}^{N_2}\frac{1}{p^{1/2-iE}}=e^{2\pi i\bar{N}(E)}
\left (\sum_{k=1}^{\infty}\sum_{n=1}^{\infty}
\frac{1}{n^{1/2+iE}}\frac{\mu(k)}{k^{1/2-iE}}
\right )_{T_H-\ln N_2<t<T_H-\ln N_1},
\end{equation}
where the symbol $t_1<t<t_2$ means that one has to expand the expression in the
brackets into the Fourier series on $\exp (-iE t)$ and to take into account only
terms with period $t_1<t<t_2$. It is important to note that if
$\ln N_1$ is of the order of $T_H$ the summation in the above formula includes only 
small number of terms. But 
\begin{equation}
\sum_{k=1}^{\infty}\frac{\mu(k)}{k^{1/2-iE}}=\frac{1}{\zeta(1/2-iE)},
\end{equation}  
and these sum can be rewritten as follows
\begin{equation}
\sum_{N_1<p<N_2}\frac{1}{p^{1/2-iE}}=e^{2\pi i\bar{N}(E)}
\left (\frac{\zeta(1/2+iE)}{\zeta(1/2-iE)}\right )_{T_H-\ln N_2<t<T_H-\ln N_1}.
\end{equation}
Using the prime number theorem the main  part of the sum of large primes in 
the trace formula  for the density  of the Riemann zero can be rewritten in 
the form
\begin{equation}
  \sum_{p=N_1}^{N_2}\frac{\ln p}{p^{1/2-iE}}=
  \frac{N_2-N_1}{\sum_{N_1<p<N_2}1}\sum_{N_1<p<N_2}p^{-1/2+iE}.
\end{equation}
Let us  use the  inclusion-exclusion principle for the denominator
\begin{equation}
\sum_{p=N_1}^{N_2}1=(N_2-N_1)(1-\sum_p\frac{1}{p}
+\sum_{p_1,p_2}\frac{1}{p_1p_2}+\ldots)=(N_2-N_1)\prod_p (1-\frac{1}{p}).
\end{equation}
Therefore
\begin{equation}
d_2(E)=-\frac{e^{2\pi i\bar{N}(E)}}{2\pi}\left (
\prod_p\frac{(1-p^{-1/2+iE})}{(1-p^{-1/2-iE)})(1-p^{-1})}
\right )_{0<t<T_H-\ln N}+c.c.,
\end{equation}
and the density of Riemann zeros takes the form 
\begin{eqnarray}
d(E)=&&-\frac{1}{2\pi} \sum_{p<N}
\ln p\sum_{n=1}^{\infty}\frac{1}{p^{n/2}}e^{iEn\ln p} 
\\
&& -\frac{e^{2\pi i\bar{N}(E)}}{2\pi}\left (
\prod_p\frac{(1-p^{-1/2+iE})}{(1-p^{-1/2-iE)})(1-p^{-1})}
\right )_{0<t<T_H-\ln N}+c.c.
\nonumber
\end{eqnarray}
This formula can be considered as a Riemann-Siegel look-like ressumation
of the trace formula. It expresses the (unknown) sum over large primes 
through small ones.  

In particular this  formula is useful for the computation of  statistics 
of Riemann zeros. Direct application of the definition (\ref{34}) gives
\begin{eqnarray}
R_2(\epsilon)&=&R_2^{(diag)}
\\
&+&\frac{e^{2\pi i \bar{d} \epsilon}}{4\pi^2}
<\prod_p \frac{(1-p^{-1/2+i(E+\epsilon_1)})(1-p^{-1/2-i(E+\epsilon_2)})}
  {(1-p^{-1/2-i(E+\epsilon_1)})(1-p^{-1/2+i(E+\epsilon_2)})(1-p^{-1})^2}>.
\nonumber
\end{eqnarray}
Here $R_2^{(diag)}$ is the same as in Eq.~(\ref{69})  except that the
summation includes primes up to $N$. Because $N$ is assumed to be large and
the sum in (\ref{69}) converges conditionally at real $\epsilon$ one can go
to the limit $N\rightarrow \infty$ and Eqs.~(\ref{70})-(\ref{72}) will be
recovered. The second term,  corresponding to $R_2^{(osc)}$,
comes from the summation over
large primes and it has the form as in Eq.~(\ref{126}) (but with correct
pre-factor). Repeating the same calculation as in Section \ref{offriemann} and
taking into account that the final product now converges one will get 
Eq.~(\ref{132}) with the correct value (\ref{135}) of the pre-factor.

\section{Conclusion}

The above-discussed methods demonstrate how, in
principle, the existence of the trace formula and certain natural
conjectures about the distribution of periodic orbits (or primes) combine together
to produce universal local statistics. For systems without
the time-reversal invariance the assumption that  periods for
all (or at least for most) periodic orbits  are
non-commensurable leads to the GUE statistics (for 2-point
correlation function).

In all cases the main formulas may be written in the following forms 
\begin{equation}
R_2(\epsilon)=R_2^{(diag)}(\epsilon)+R_2^{(off)}(\epsilon),
\end{equation}
where the diagonal part 
\begin{equation}
R_2^{(diag)}(\epsilon)=-\frac{1}{4\pi^2}
  \frac{\partial^2}{\partial \epsilon^2}
  \ln (|Z_{cl}(\epsilon)|^2\Phi^{(diag)}(\epsilon)),
\end{equation}  
and the off-diagonal part 
\begin{equation}
R_2^{(off)}(\epsilon)=\frac{1}{4\pi^2}e^{2\pi i \bar{d}\epsilon}
|\frac{1}{\gamma}Z_{cl}(\epsilon)|^2\Phi^{(off)}(\epsilon).
\end{equation}
Here $Z_{cl}(\epsilon)$ is a classical zeta function  and
$\Phi^{(diag)}(\epsilon)$ and $\Phi^{(off)}(\epsilon)$ are certain convergent
products over periodic orbits.  

For dynamical systems
\begin{equation}
Z_{cl}(\epsilon)=\prod_p(1-\exp(i\epsilon T_p)/|\Lambda_p|)^{-1}.
\end{equation}

For the Riemann case
\begin{equation}
Z_{cl}(\epsilon)=\zeta(1-iE).
\end{equation}
In all cases $Z_{cl}(\epsilon)\rightarrow \gamma/\epsilon$ when
$\epsilon\rightarrow 0$ and $\Phi^{(off)}(0)=1$ which ensures the GUE value
for small $\epsilon$.

The close relation between diagonal and off-diagonal  terms (first observed
for disordered systems by Andreev and Altshuler \cite{Altshuler}) suggests the 
existence of a certain unified principle. 
The best candidate for it is  the `unitarity' property of
the trace formula, namely, that the distribution of periodic orbits should be such
that the corresponding eigenvalues will be  real. In some sense certain long-period
orbits are connected to the short ones and the investigation of this
connection may clarify the origin of universal spectral statistics. 

The interesting question remains what conjectures about periodic orbits are necessary
to obtain correlation functions for systems with time-reversal invariance
where almost all periodic orbits appear in pairs with exactly the same action.

{\bf Acknowledges} The author is greatly indebted to Jon Keating in
collaboration with whom many results of this paper has been derived.

\end{document}